\documentclass[10pt,final,journal,a4paper,oneside,twocolumn]{IEEEtran}

\usepackage{subfigure}
\usepackage{srcltx}
\usepackage{psfrag}
\usepackage{epsfig}
\usepackage{color}
\usepackage[tbtags,sumlimits,nointlimits,reqno]{amsmath}
\usepackage{amssymb}
\usepackage{showkeys}   %show the labels, cite...
\usepackage{color}
\usepackage{float}
\usepackage{mathtools, cuted}
\usepackage[makeroom]{cancel}
\usepackage{enumerate}
\begin{document}

\title{Integrated Generalized Sheet Transition Conditions (GSTCs) in a Yee-Cell based Finite-Difference Time-Domain (FDTD) Simulation of Electromagnetic Metasurfaces}

\author{%
       Tom. J. Smy, Scott Stewart and Shulabh~Gupta
\thanks{T. J. Smy, Scott Stewart and S.~Gupta, are with the Department of Electronics, Carleton University, Ottawa, Ontario, Canada. Email: mohamem@doe.carleton.ca}
}
 %The paper headers
\markboth{MANUSCRIPT DRAFT}
{Shell \MakeLowercase{\textit{et al.}}: Bare Demo of IEEEtran.cls for Journals}

% make the title area
\maketitle
\begin{abstract}
A finite-difference time-domain (FDTD) simulation of broadband electromagnetic metasurfaces based on direct incorporation of Generalized Sheet Transition Conditions (GSTCs) inside a conventional Yee-cell region has been proposed, for arbitrary wave excitations. This is achieved by inserting a zero thickness metasurface inside bulk nodes of the Yee-cell region, giving rise to three distinct cell configurations - Symmetric Cell (SC), Asymmetric Cell (AC) and Tight Asymmetric Cell (TAC). In addition, the metasurface is modelled using electric and magnetic surface susceptibilities exhibiting a broadband Lorentzian response. As a result, the proposed model guarantees a physical and causal response from the metasurface. Several full-wave results are shown, and compared with analytical Fourier propagation methods showing excellent results, for both 1D and 2D fields simulations. It is found that the TAC provides the fastest convergence among the three methods with minimum error.  
\end{abstract}


%\tableofcontents

\section{Introduction}

%\begin{enumerate}
%\item Brief review of metasurfaces \cite{Metasurface_Review}. 
%\item Generalized Sheet Transition conditions (GSTCs).
%\begin{itemize}
%	\item Field equations 
%	\item Frequency domain versus time domain (motivation for time domain)
%	\item Discussion of $\chi$ (constant and generalized a sum of Lorentzians). Causality?
%	\item Matched surfaces
%\end{itemize}
%\item Yee Cell and FDTD
%
%
%\begin{itemize}
%	\item Basic Yee Cell description
%	\item GSTC using explicit surface (pros and cons)
%	\item Three possible surface cells
%\end{itemize}
%\item Contribution in this work.
%	\begin{enumerate}
%	\item Direct incorporation into GSTCs to create an implicit Yee Cell with GTSCs
%	\item Stability and Accuracy 
%		\begin{itemize}
%			\item Range of constant $\chi$
%			\item Lorentzian accuracy -- reflection
%			\item Comparison with explicit surface
%		\end{itemize}
%	\end{enumerate}
%\item Structure of the Paper.	
%\end{enumerate}

Electromagnetic (EM) metasurfaces are two-dimensional equivalents of volumetric metamaterials, and are composed of 2D arrays of sub-wavelength scatterers. By engineering these scatterers across the surface, various interesting wave-shaping transformations can be achieved for various applications such as generalized refraction, holography, polarization control, imaging and cloaking, to name a few \cite{GeneralizedRefraction}\cite{meta3}. Metasurfaces achieve such wave transformations as a result of complex interplay between the electric and magnetic dipolar moments generated by the scatterers, which is sometimes also referred to as a Huygens' configuration \cite{Elliptical_DMS}\cite{Kivshar_Alldielectric}. A convenient implementation of such metasurfaces is using all-dielectric resonators, which naturally produce the electric and magnetic dipoles moments, and when properly designed, provide zero backscattering, resulting in a perfect transmission \cite{Kerker_Scattering}\cite{AllDieelctricMTMS}\cite{Grbic_Metasurfaces}.

A recently growing area of interest is reconfigurable and time-varying metasurfaces, where the constitutive parameters (surface susceptibilities, $\chi_\text{e,m}$) of the metasurfaces are real-time tunable. A more general description of such dynamic metasurfaces is a space-time modulated metasurface, where the surface susceptibilities are both a function of space and time, resulting in a travelling-type perturbation on the metasurface. They are the 2D equivalents of general space-time modulated mediums \cite{TamirST}\cite{OlinerST}, which have found important applications in parametric amplifiers and acousto-optic spectrum analyzers, \cite{TamirAcoustoDiffraction}\cite{Goodman_Fourier_Optics}\cite{Saleh_Teich_FP}, for instance. Space-time modulation has led to various exotic effects such as harmonic generation and non-reciprocity \cite{Caloz_ST_Modulated}\cite{Mechanicl_ST_Modulation}, that has also been recently explored using metasurfaces \cite{STGradMetasurface}\cite{ShaltoutSTMetasurface} for advanced wave-shaping applications. Their attractive features lies in achieving non-reciprocity using purely non-magnetic materials, which has important practical benefits in engineering systems, related to high frequency operation and no requirement of a magnetic bias.

The EM modelling of metasurfaces with such advanced wave manipulations necessitates a need for efficient time-domain simulation of these structures. While practical metasurfaces are sub-wavelength in thickness ($\delta \ll \lambda_0$), they can be efficiently modelled as space-discontinuities, described using electric and magnetic surface susceptibilities, i.e. $\tilde{\chi}_\text{ee}(\omega)$ and $\tilde{\chi}_\text{mm}(\omega)$, respectively. They thus model practical metasurfaces as zero thickness structures, thereby transforming them into a single-interface problem. Such space-discontinuities can be rigorously modelled using Generalized Sheet Transition Conditions (GSTCs). While various numerical approaches have been recently presented, where the GSTC conditions are incorporated in the finite-difference formulation in frequency domain to accurately analyze the transmitted and reflected fields of a general zero-thickness metasurface \cite{CalozFDTD}\cite{FEM_Metasurafce}, very little work has been done in time-domain modelling of metasurfaces \cite{Smy_Metasurface_Linear}. In this work, a rigorous Finite-Difference Time-Domain (FDTD) method is developed, where the GSTCs are directly integrated into the FDTD Yee-cells, similar to that in previous works on frequency-domain models. Compared to the explicit FDTD model presented in \cite{Smy_Metasurface_Linear}, where the metasurface is treated as a boundary of a given simulation domain, the proposed method treats the metasurface as an EM scattering entity, and is able to process arbitrary broadband excitations. 

In contrast to frequency-domain modelling, the time-domain modelling of EM metasurfaces explicitly requires causality considerations for accurate modelling of practical metasurface responses. Since metasurfaces are constructed using sub-wavelength resonators, they are operated around the resonant frequencies where the EM waves have maximum interaction with the metasurface. Consequently, these metasurfaces are naturally very \emph{dispersive}, i.e. $\tilde{\chi}_\text{e,m}(\omega) \ne \text{const.}$ The geometrical shapes of the constituting scatterers are primarily responsible for their resonant behaviour, in spite of their design generally based on non-dispersive materials (typically metals and dielectrics). This operation of the metasurface in a dispersive (and thus broadband) regime, demands a physical description of these resonators consistent with the \emph{causality requirements}. This in turn, requires a causal description of the equivalent surface $\chi_\text{e,m}$ of the metasurface, in frequency (or time domain), i.e. $\tilde{\chi}_\text{e,m} = \tilde{\chi}_\text{e,m}(\omega)$ or $\chi_\text{e,m} = \chi_\text{e,m}(t)$. This requirement is also critical in the accurate time-domain modelling of general space-time modulated metasurfaces, where new spectral frequency components are generated. This subsequently further necessitates a complete description of the surface $\chi_\text{e,m}$ encompassing these frequencies as well, in addition to the bandwidth of the input excitation. 

In this context, various strategies for integrating GSTCs in a conventional FDTD Yee-cell is proposed and compared here,  assuming Lorentzian surface susceptibilities, which are naturally causal and rigorously capture the fundamentally dispersive nature of typical EM metasurfaces. The paper is structured as follows. Section II presents the GSTC formulation of zero thickness metasurfaces, and establishes analytical models for specific cases, for benchmarking purposes. Section III proposes three possible different Yee-cell configurations where GSTCs can be incorporated in conventional cells. Section IV shows several simulated results corresponding to these configurations, providing detailed comparisons between them. Finally Sec. V provides concluding discussions and remarks on the applicability of the Lorentzian surface susceptibilities and conclusions are provided in Sec. VI. Some important field derivations of the proposed Yee-cell configurations are also provided in the Appendix.

\section{Metasurface Description}

\subsection{Generalized Sheet Transition Conditions (GSTCs)}

\begin{figure}[tbp]
\begin{center}
\psfrag{A}[c][c][0.8]{$H_0$}
\psfrag{B}[r][c][0.8]{$H_r$}
\psfrag{D}[c][c][0.8]{$E_0$}
\psfrag{F}[c][c][0.8]{$E_t$}
\psfrag{E}[c][c][0.8]{$H_t$}
\psfrag{G}[c][c][0.8]{$E_r$}
\psfrag{x}[c][c][0.8]{$x$}
\psfrag{z}[c][c][0.8]{$z$}
\psfrag{J}[c][c][0.8]{$z = 0$}
\psfrag{a}[c][c][0.8]{$\mathbf{m_x}~[\tilde{\chi}_\text{mm}]$}
\psfrag{b}[c][c][0.8]{$\mathbf{p_y}~[\tilde{\chi}_\text{ee}]$}
\psfrag{c}[c][c][0.8]{$\mathbf{k_z}$}
\includegraphics[width=0.8\columnwidth]{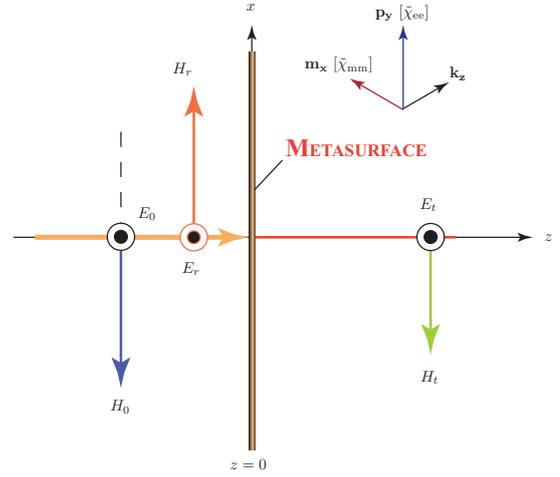}
\caption{Typical configuration of a zero-thickness uniform Huygen's metasurface, located at $z=0$, consisting of orthogonal electric ($\mathbf{p}$) and magnetic ($\mathbf{m}$) dipole moments, excited with a normally incident plane-wave resulting in reflected and transmitted fields governed by \eqref{Eq:GSTC}. For simplicity, normal polarization is assumed without any variation of the fields along the $y-$axis.}\label{Fig:Problem}
\end{center}
\end{figure}

A zero thickness metasurface, such as the one in Fig.~\ref{Fig:Problem}, is a space-discontinuity. The rigorous modelling of such discontinuities based on Generalized Sheet Transition Conditions (GSTCs) were developed by Idemen in \cite{IdemenDiscont}, which were later applied to metasurface problems in \cite{KuesterGSTC}. The modified Maxwell-Faraday and Maxwell-Ampere equations  can be written in the time-domain as,

\begin{subequations}\label{Eq:GSTC}
\begin{equation}
\hat{\mathbf{z}}\times\Delta \mathbf{H}(x,t) = \frac{d\mathbf{P}_s(x,t)}{dt}\label{Eq:GSTCa}
\end{equation}
\begin{equation}
\Delta \mathbf{E}(x,t)\times \hat{\mathbf{z}} = \mu_0\frac{d\mathbf{M}_s(x,t)}{dt},\label{Eq:GSTCb}
\end{equation}
\end{subequations}

\noindent where $\Delta \psi$ represents the differences between the fields on the two sides of the metasurface for all the vector component of the field $\psi$, i.e. $\mathbf{H}$ or $\mathbf{E}$ fields. The other terms $\mathbf{P}_\text{s}$ and $\mathbf{M}_\text{s}$ represent the electric and magnetic surface polarization densities, \emph{in the plane of the metasurface}, which depend on the total average fields around the metasurface \cite{Metasurface_Synthesis_Caloz}. They are defined by,	

\begin{subequations}
\begin{equation}
\tilde{\mathbf{P}_\text{s}}(\omega) = \epsilon_0 \tilde \chi_\text{ee}(\omega) \tilde{\mathbf{E}_\text{s}}(\omega)
\end{equation}
\begin{equation}
\tilde{\mathbf{M}_\text{s}}(\omega) = \tilde \chi_\text{mm}(\omega) \tilde{\mathbf{H}_\text{s}}(\omega) 
\end{equation}
\end{subequations}

\noindent where $\tilde{\chi}_\text{ee}$ and $\tilde{\chi}_\text{mm}$ are the frequency dependent electric and magnetic susceptibilities and $\tilde{\mathbf{E}}_\text{s}$ and $\tilde{\mathbf{H}}_\text{s}$ the average EM fields at the surface. The EM coupling related to the bi-anisotropic term is assumed zero here, for simplicity. Furthermore, the surface susceptibilities $\tilde{\chi}_\text{ee}$ and $\tilde{\chi}_\text{mm}$ are also treated a scalars, as opposed to their most general tensorial forms to account for more general wave transformations. 

\subsection{Surface Polarization Densities}

A primary concern in modeling the metasurface response is a physical representation of the surface polarizations consistent with \emph{causality}. The metasurface units cell will have a number of natural resonances and this response must be captured for the correct broadband response to be predicted. These resonances are naturally modeled by Lorentzian functions and a summation of correctly parameterized Lorentzian's is an appropriate model of the surface, chosen in this work, i.e.

\begin{subequations}	
\begin{equation}
\tilde \chi_\text{ee}(\omega)  = \sum_{n=0}^N \frac{\epsilon_0 \omega_{ep, n}^2}{(\omega_{e0,n}^2 - \omega^2) + j\alpha_{e, n}\omega}  
\end{equation}
\begin{equation}
\tilde \chi_\text{mm}(\omega)  = \sum_{n=0}^N \frac{\epsilon_0 \omega_{mp, n}^2}{(\omega_{m0,n}^2 - \omega^2) + j\alpha_{m, n}\omega}  \end{equation}
\end{subequations}

\noindent A key consideration in the use of Lorentzians is that they represent a physical process and therefore are implicitly causal. In addition, they naturally take into account the dispersive effects of the metasurface, which have practical importance in the EM interaction of metasurfaces with broadband excitations.

While a typical metasurface requires several Lorentzian contributions to accurately model broadband surface susceptibilities, such as for all-dielectric unit cells \cite{Smy_Metasurface_Linear}, here we assume a single resonant Lorentz response for the electric and magnetic polarizations, for the sake of clarity and simpler forthcoming analytical expressions, so that 

\begin{subequations}	\label{Eq:lor}
\begin{equation}
\tilde{\mathbf{P}_\text{s}}(\omega)  = \frac{\epsilon_0 \omega_{ep}^2}{(\omega_{e0}^2 - \omega^2) + i\alpha_e\omega} \tilde{\mathbf{E}_\text{s}}(\omega)  
\end{equation}
\begin{equation}
\tilde{\mathbf{M}_\text{s}}(\omega)  = \frac{\omega_{mp}^2}{(\omega_{m0}^2 - \omega^2) + i\alpha_m\omega} \tilde{\mathbf{H}_\text{s}}(\omega), 
\end{equation}
\end{subequations}

\noindent where $\omega_p$, $\omega_0$ and $\alpha$ are the plasma frequency, resonant frequency and the loss-factor of the oscillator, respectively, and subscript $e$ and $m$ denote electric and magnetic quantities. It should be noted that the use of a single Lorentzian here is strictly for simplicity and the proposed method below can easily be extended for multiple resonance contributions, due to linear field superposition. 

To formulate a time-domain Yee surface cell we need a time-domain representation for the surface polarizations and must transform (\ref{Eq:lor}) to the time domain, using the inverse Fourier transform, leading to

\begin{subequations}
\begin{equation}
	\frac{d^2 \mathbf{P}_\text{s}}{dt^2} + \alpha_e \frac{d\mathbf{P}_\text{s}}{dt} + \omega_{e0}^2 \mathbf{P}_\text{s} = \epsilon_0 \omega_{ep}^2 \mathbf{E}_\text{s}
\end{equation}
\begin{equation}	
\frac{d^2 \mathbf{M}_\text{s}}{dt^2} + \alpha_m \frac{d\mathbf{P}_\text{s}}{dt} + \omega_{m0}^2 \mathbf{P}_\text{s} = \omega_{mp}^2 \mathbf{H}_\text{s}
\end{equation}
\end{subequations}

\noindent It is convenient to formulate these two equations as a $1^\text{st}$ order system using two variables. For example, for the case of Fig.~\ref{Fig:Problem}, defining $P_s$ and $\omega_{e0} P_s' = dP_s/dt$ along the $y-$axis, we obtain,

\begin{align*}
	\left[
	\begin{array}{cc}
		1 & 0 \\
		0 & 1 
	\end{array}
	\right]
	\frac{d}{dt} \left[
	\begin{array}{c}
		P_s'\\
		P_s
	\end{array}
	\right] &+
	\left[
	\begin{array}{cc}
		\alpha_e  & \omega_{e0} \\
		-\omega_{e0}  & 0 
		\end{array}		 
		\right]
		\left[
		\begin{array}{c}
			P_s'\\
			P_s
		\end{array}
		\right] = \\
		&\left[
		\begin{array}{c}
			\epsilon_0\omega_{ep}^2/\omega_{e0}\\
			0
		\end{array}
		\right] E_s		 
\end{align*}

\noindent allowing us to write the first equation as, 

\begin{equation}
	\label{Eq:Pt}
	[\mathbf{C_p}] \frac{d [\mathbf{P_s}]}{dt} + [\mathbf{G_p}] [\mathbf{P_s}] = [\mathbf{F_p}] E_s , \; [\mathbf{P_s}] = [P'_s\; P_s]^T
\end{equation}

\noindent In a similar manner we can define $M_s$ and $\omega_{m0} M_s' = dM_s/dt$ and obtain,

\begin{equation}
	\label{Eq:Mt}
	[\mathbf{C_m}] \frac{d [\mathbf{M_s}]}{dt} + [\mathbf{G_m}] [\mathbf{M_s}] =[ \mathbf{F_m}] H_s , \; [\mathbf{M_s}] = [M'_s\;M_s]^T.
\end{equation}

%It should be noted that although for narrowband simulations a constant susceptibility might be appropriate, such a representation of the surface for broadband time domain simulations will create stability problems due to the non-physical nature of a such a characterization.

\subsection{Fourier Transform Solution for Normally Incident Plane-wave}
\label{sec:FT}

Let us first consider a specific case of a linear and uniform metasurface, which is excited with a normally incident pulsed plane-wave. It represents a simple case, whose time-domain transmitted and reflected fields can be obtained using standard Fourier propagation method \cite{Agrawal_NLFO_1980}, which provides a baseline for comparison to the Yee cell simulations. Now, consider a Huygens' metasurface illuminated with a normally incident plane-wave, as shown in Fig.~\ref{Fig:Problem}. For simplicity, but without loss of generality, the problem is assumed to be 2D, where all $y-$variations are assumed to be zero. For normal incidence, the input, transmitted and reflected plane-waves are given by

\begin{subequations}	\label{eq:fields}
\begin{align}
\mathbf{E_0}(z,t) & = E_0(t) e^{j(\omega t -kz)}~\mathbf{\hat{y}} ,~\mathbf{H_0}= \frac{ \mathbf{\hat{z}}  \times \mathbf{E_0}}{\eta_0}\\
\mathbf{E_t}(z,t) & = E_t(t) e^{j(\omega t -kz)}~\mathbf{\hat{y}} ,~\mathbf{H_t}= \frac{ \mathbf{\hat{z}}  \times \mathbf{E_t}}{\eta_0} \\
\mathbf{E_r}(z,t) & = E_r(t) e^{j(\omega t +kz)}~\mathbf{\hat{y}} ,~\mathbf{H_r}= \frac{\mathbf{E_r} \times \mathbf{\hat{z}}}{\eta_0}
\end{align}
\end{subequations}

\noindent where $\eta_0$ is the free-space impedance. Let us assume a monochromatic excitation first with a frequency $\omega$. Substituting the above fields into \eqref{Eq:GSTC} using,

\begin{align}
\tilde{\mathbf{P}}_{s}(\omega) &= \epsilon_0\tilde{\chi}_\text{ee}(\omega) \frac{\tilde{\mathbf{E}}_t(\omega) + \tilde{\mathbf{E}}_r(\omega) + \tilde{\mathbf{E}}_0(\omega)}{2}, \notag\\
\tilde{\mathbf{M}}_{s}(\omega) &= \tilde{\chi}_\text{mm}(\omega) \frac{\tilde{\mathbf{H}}_t(\omega) + \tilde{\mathbf{H}}_r(\omega) + \tilde{\mathbf{H}}_0(\omega)}{2}
\end{align}

\noindent and Fourier transforming, leads to,

%\begin{subequations}\label{Eq:TDGSTC}
%\begin{equation}
%(-H_t - H_r + H_0) = \frac{\epsilon_0\tilde{\chi}_\text{ee}(\omega)}{2}\frac{d}{dt}(E_t + E_r + E_0)
%\end{equation}
%\begin{equation}
%(E_t - E_r - E_0) = \frac{\mu_0\tilde{\chi}_\text{mm}(\omega)}{2}\frac{d}{dt}(-H_t + H_r -H_0).
%\end{equation}
%\end{subequations}
%
%\noindent This can then be transformed into the frequency domain resulting in:

\begin{subequations}\label{Eq:TDGSTCF}
\begin{equation}
	\label{Eq:TDGSTCFa}
(-\tilde E_t - \tilde E_r + \tilde E_0) = j \omega \frac{\epsilon_0\eta_0\tilde{\chi}_\text{ee}}{2}(\tilde E_t + \tilde E_r + \tilde E_0)
\end{equation}
\begin{equation}
	\label{Eq:TDGSTCFb}
(\tilde E_t - \tilde E_r - \tilde E_0) = j \omega \frac{\mu_0\tilde{\chi}_\text{mm}}{2\eta_0}(-\tilde E_t + \tilde E_r - \tilde E_0).
\end{equation}
\end{subequations}

% \begin{align*}
% (-\tilde E_t - \tilde E_r + \tilde E_0) &= \tilde \Gamma_e(\omega) (\tilde E_t + \tilde E_r + \tilde E_0)\\
% (\tilde E_t - \tilde E_r - \tilde E_0) &= \tilde \Gamma_h(\omega) (-\tilde E_t + \tilde E_r - \tilde E_0)
% \end{align*}

\noindent Equations (\ref{Eq:TDGSTCFa}) and (\ref{Eq:TDGSTCFb}) can further be placed into a matrix form by defining,

\begin{align*}
	[\mathbf{\tilde H}] =  
	\left[
	\begin{array}{cc}
		-\tilde \Gamma_e-1 & -\tilde \Gamma_e-1\\
		\tilde \Gamma_h+1 & -\tilde \Gamma_h-1\\
		\end{array}	
		\right],~\text{and}~
		[\mathbf{\tilde F}] = 
		\left[
		\begin{array}{c}
			\tilde \Gamma_e-1\\
			-\tilde \Gamma_h+1
		\end{array}
		\right]
	\end{align*}
	
\noindent \noindent where $\Gamma_e = j \omega \epsilon_0\eta_0\tilde{\chi}_\text{ee}/2$ and $\tilde \Gamma_h = j \omega \mu_0\tilde{\chi}_\text{mm}/2\eta_0$, resulting in,

\begin{align*}
 	& 
	[ \mathbf{\tilde H}]\left[
	 \begin{array}{c}
		 \tilde E_t\\
		 \tilde E_r
	 \end{array}
	 \right]
	 =
	 [\mathbf{\tilde F}] \tilde E_0
\end{align*}

%\begin{align*}
%	\tilde \Gamma_e = j \omega \frac{\epsilon_0\eta_0\tilde{\chi}_\text{ee}}{2}, \quad  
%	\tilde \Gamma_h = j \omega \frac{\mu_0\tilde{\chi}_\text{mm}}{2\eta_0}
%\end{align*}
%
%and

\noindent Using this equation after specifying $\tilde E_0(\omega) = \mathcal{F}\{E_0(t, 0_-)\}$, we can then determine $E_t(t)$ and $E_r(t)$ using an inverse Fourier transform, i.e.

\begin{align}\label{Eq:FT}
	\left[
	 \begin{array}{c}
		 E_t(t, 0_+)\\
		 E_r(t, 0_-)
	 \end{array}
	 \right]
= \mathcal{F}^{-1} 
 \left \{
 [\mathbf{\tilde H}]^{-1}
[\mathbf{\tilde F}] \tilde E_0
 \right \}
\end{align}

\noindent Finally, the instantaneous propagating fields as a function of $z$ are then obtained using (\ref{eq:fields}).

\section{Yee Cell Formulation}
\begin{figure}[tbp]
\begin{center}
\psfrag{x}[c][c][0.8]{$x$}
\psfrag{y}[c][c][0.8]{$y$}
\psfrag{G}[c][c][0.8]{$\Delta$}
\psfrag{B}[l][c][0.8]{$E_y$-nodes}
\psfrag{C}[l][c][0.8]{$H_x$-nodes}
\psfrag{D}[l][c][0.8]{$H_z$-nodes}
\psfrag{E}[l][c][0.8]{surface update nodes}
\psfrag{F}[l][c][0.8]{irregular update nodes}
\psfrag{a}[c][c][0.6]{$(i,j)$}
\psfrag{m}[c][c][0.6]{$s_-$}
\psfrag{n}[c][c][0.6]{$s_+$}
\psfrag{o}[c][c][0.6]{$k+1/2$}
\psfrag{p}[c][c][0.6]{$k$}
\psfrag{q}[c][c][0.6]{$k-1$}
\psfrag{r}[c][c][0.6]{$k-3/2$}
\psfrag{b}[c][c][0.6]{$i-1$}
\psfrag{c}[c][c][0.6]{$i-1/2$}
\psfrag{d}[c][c][0.6]{$i$}
\psfrag{e}[c][c][0.6]{$i+1/2$}
\psfrag{f}[c][c][0.6]{$i+1$}
\psfrag{g}[r][c][0.6]{$j-1$}
\psfrag{h}[r][c][0.6]{$j-1/2$}
\psfrag{i}[r][c][0.6]{$j$}
\psfrag{j}[r][c][0.6]{$j+1/2$}
\psfrag{k}[r][c][0.6]{$j+1$}
\includegraphics[width=\columnwidth]{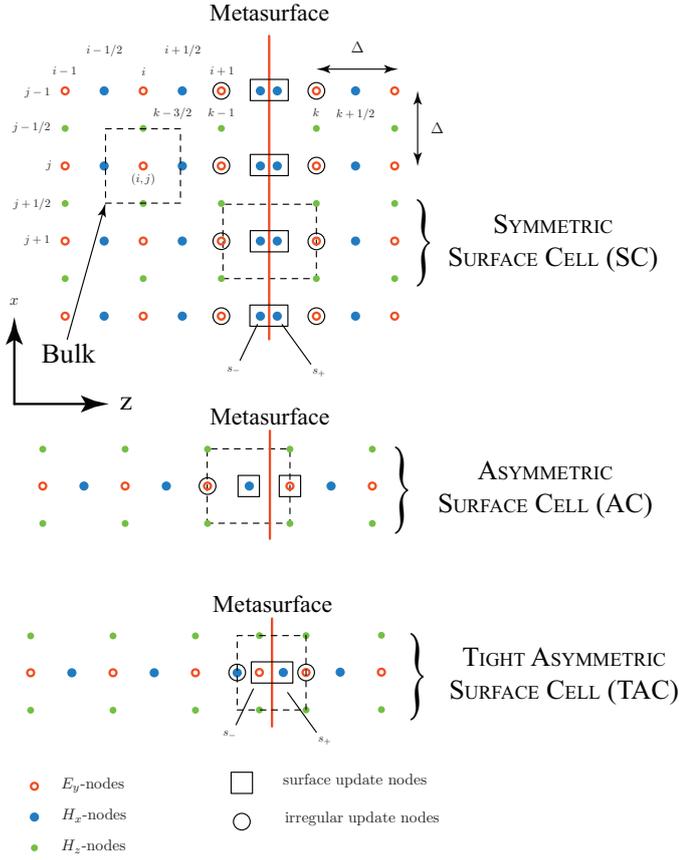}
\caption{Three different Yee-cell configurations integrating a zero thickness metasurface in bulk nodes, investigated in this work.}\label{Fig:YeeCell}
\end{center}
\end{figure}

%\subsection{Bulk Yee-Cell Formulations}
In this section, we will describe the integration of GSTCs into bulk Yee-cells, using three possible configurations. The formulation will be developed for 2D propagation with propagation in the $x-z$ plane, however, the proposed approach can be straightforwardly extended into a full 3D simulation. The total simulation region consists of two types of nodes: bulk nodes for modelling the reflection and transmission region following Maxwell's equations, and surface-nodes modelling the zero thickness metasurface following the GSTCs. For a bulk 2D Yee-cell defined for propagation on the $x-z$ plane (see Fig.~\ref{Fig:YeeCell}), the basic equations for the electric and magnetic fields, are given by conventional update equations,

\begin{align*}
	&{H_z|_{i,j+1/2}^{n+1/2}} = H_z|_{i,j+1/2}^{n-1/2} - \frac{\Delta t}{\mu}\left(\frac{E_y|_{i,j+1}^n - E_y|_{i,j}^n}{\Delta}\right)\\
	&{H_x|_{i+1/2,j}^{n+1/2}} = H_x|_{i+1/2,j}^{n-1/2} + \frac{\Delta t}{\mu}\left(\frac{E_y|_{i+1,j}^n - E_y|_{i,j}^n}{\Delta}\right)\\
	&{E_y|_{i,j}^{n+1}} = E_y|_{i,j}^{n} + \\
	&\frac{\Delta t}{\epsilon}\left(\frac{H_x|_{i+1/2,j}^{n+1/2} -  H_x|_{i-1/2,j}^{n+1/2}}{\Delta} - 	\frac{H_z|_{i,j+1/2}^{n+1/2} - H_z|_{i,j-1/2}^{n+1/2}}{\Delta}\right)
\end{align*}

\noindent where the subscript is the spatial position (with $i$ and $j$ representing node positions in $z$ and $x$ respectively) and the superscript denotes the time-step. A field evolution is then obtained by stepping in time using the following procedure:

\begin{enumerate}
	\item Update the H's at $n+1/2$ ($t_{n+1/2}$= $t_n$ + $\Delta t/2$) using previous E's and the boundary conditions.
	\item Update E's at $n+1$ ($t_{n+1}$ = $t_n$ + $\Delta t$) using the H's calculated at $n+1/2$.
\end{enumerate}

\noindent A key feature of this procedure is that the current fields being determined are found using only previously calculated fields. For example $E^{n+1}$'s are determined from the $H^{n+1/2}$'s. Therefore the method is a strict explicit time marching method. As an explicit method the spatial step, $\Delta$, and the time step, $\Delta t $, should always satisfy the Courant condition $u \sqrt{2} \Delta t \le \Delta $ (where $u$ is the local speed of light) to guarantee stability \cite{taflove2000computational}.

Next is the metasurface region. When forming the update equations for the metasurface cells, the nature of the GSTC equations (\ref{Eq:GSTC}) require that the $H_x|^{n+1/2}$ and the $E_y|^{n+1}$ fields are coupled and cannot be solved in a simple sequential manner. This is due to the polarizations being naturally solved at the time point associated with their fields and thus we solve for $M_s|^{n+1/2}$ and $P_s|^{n+1}$. Therefore, it is needed to formulate a self-consistent solution to the unknowns present in each cell and then update all fields after a complete time step. 

\subsection{Yee-Cell Configurations}

To incorporate the GSTCs modelling a zero thickness metasurface, inside a bulk Yee-cell region with minimal disruption, there are three different possibilities. The three following surface cells we wish to consider are shown in Fig. \ref{Fig:YeeCell}:

\begin{enumerate}
\item {\bf Symmetrical Surface Cell (SC):} This cell is identical in form to the cell used in \cite{Smy_Metasurface_Linear} for formulating an explicit surface as an internal boundary condition. The surface is inserted midway between two electric field nodes ($E_y|_{k-1}$ and $E_y|_k$) and the $H_x$ node present at the position $k$-1/2 is split into two nodes on either side of the surface denoted as $H_x|_{s^-}$ and $H_x|_{s^+}$.

\item {\bf Asymmetrical Surface Cell (AC):} The second cell is formed by simply inserting the surface midway between an $H_x$ node and $E_y$ node. This requires no new nodes to be defined but produces an asymmetrical cell structure. 

\item {\bf Tight Asymmetrical Surface Cell (TAC):} The third structure is similar to the AC cell but inserts new nodes either side of the surface. On the left side a new electric field node is inserted ($E_y|_{s^-}$) and on the right a magnetic field node is inserted ($H_x|_{s^+}$).
\end{enumerate}

\subsection{Yee-Cell Unknowns}

The need for a self-consistent solution for the surface cell requires a clear identification of the variables defined at and near the surface. Although some of the nodes are essentially bulk nodes (see for example the $E_y|_k$ node in the SC cell) a special update for this node is needed as the $H_{s^+}|^{n+1/2}$ node value is not known until the surface cell is solved. Inspection of the three cells allows us to identify quantities that would be unknown prior to a surface update (assuming all bulk nodes $H_x|^{n+1/2}$, $H_z|^{n+1/2}$ and $E_y|^{n+1}$ have been updated) and we can group the unknowns in a vector $X$, given by

\begin{align*}
	% &\text{\bf Symmetric Cell}\\
	[\mathbf{X_{SC}}] & = \left[E_y|_{k-1,l}^{n+1} \; E_y|_{k,l}^{n+1}  \;  H_x|_{s^-,l}^{n+1/2} \; \right. \\ & \left. H_x|_{s^+,l}^{n+1/2} \; 
	[\mathbf{P_s}]|_{l}^{n+1/2} \; [\mathbf{M_s}]|_{l}^{n+1/2} \right]^T\\
	% &\text{\bf Asymmetric Cell}\\
	[\mathbf{X_{AC}}] & = \left[E_y|_{k,l}^{n+1} \; E_y|_{k-1,l}^{n+1}  \;  H_x|_{k-1/2,l}^{n+1/2}  \; [\mathbf{P_s}]|_{l}^{n+1/2} \; [\mathbf{M_s}]|_{l}^{n+1/2}   \right]^T\\\
	% &\text{\bf Asymmetric (tight) Cell}\\
	[\mathbf{X_{TAC}}] & = \left[E_y|_{k,l}^{n+1} \; E_y|_{s^-,l}^{n+1}  \;  H_x|_{s^+,l}^{n+1/2}  \;\right. \\ & \left.  H_x|_{k-1/2,l}^{n+1/2} \; [\mathbf{P_s}]|_{l}^{n+1/2} \; [\mathbf{M_s}]|_{l}^{n+1/2} \right]^T.
\end{align*}

\noindent Given these unknowns we now need to formulate a linear set of equations which incorporates the GSTCs of  (\ref{Eq:GSTC}) and the polarization responses given by (\ref{Eq:Pt}) and (\ref{Eq:Mt}) that can be integrated into the Yee cell algorithm.

As the update equations for the surface cells derived in the following sections are complex -- requiring the self-consistent solution of variables at two times ($n$+$1/2$ and $n$+$1$) -- to clarify the equations we have used boxed variables such as $\boxed{E_y|_{k,l}^{n+1}}$ to denote unknowns to be solved for. Other values will be known at the time of solution for the fields in the cell.

\subsection{Metasurface Field Equations}

The first GSTC equation (\ref{Eq:GSTCa}) specifies a relationship between the field across the surface ($\Delta E_y$) and the magnetic polarization. To descritize this equation we have two choices:

\begin{enumerate}[(a)]

\item As a first choice, we naturally impose the electric field across the surface at the time point $n+1$ and use central difference in time for the polarization $M_s$ (which will be solved on the half steps $n$+1/2) providing,

\begin{align*}
	% \label{Eq:1stchoice}
	\Delta E_y|_s^{n+1} &= \mu_0 \frac{\boxed{M_{s}|_l^{n+3/2}} - \boxed{M_{s}|_l^{n+1/2}}}{\Delta t}
\end{align*}

\noindent This equation is problematic, however, as $M_s|_l^{n+3/2}$ is in the \emph{future} and not part of our solution. We must therefore make an approximation and use,

\begin{align}
	\label{Eq:dE}
	\Delta E_y|_s^{n+1} &\approx  \mu_0 \frac{\boxed{M_s|_l^{n+1/2}} - M_s|_l^{n-1/2}}{\Delta t}.
\end{align}

\item The second choice is to center $\Delta E_y$ at the time step $n$, allowing us to write,

\begin{align*}
	% \label{Eq:2ndchoice}
	\Delta E_y|_s^{n} &= \mu_0 \frac{\boxed{M_s|_l^{n+1/2}} - M_s|_l^{n-1/2}}{\Delta t} 
\end{align*}

\noindent However, this choice is also not useful, as for small $\chi_\text{ee}$, the above equation become problematic. This can be seen by setting $\chi_\text{ee} = 0$  which produces,

\begin{align*}
	% \label{Eq:2ndchoice}
	\Delta E_y|_s^n &= 0.
\end{align*}

\noindent Such an equation does not provide a relationship between any of the unknowns present in $\Delta E_y|_s$ and will produce an under-determined set of cell equations. We will therefore use (\ref{Eq:dE}) in the subsequent formulation.

\end{enumerate}

The second GSTC equation (\ref{Eq:GSTCb}) relating $\Delta H_x$ to the electric polarization can be handled more straightforwardly. Using the $H_x$ values at the surface centered on the time step $n+1/2$, we have,

\begin{align}
		\label{Eq:dH}
		\Delta H_x|_{s,l}^{n+1/2} &= \frac{\boxed{P_s|_{l}^{n+1}} - P_s|_{l}^{n}}{\Delta t}.
\end{align}

\noindent Both of these equations (\ref{Eq:dE}) and (\ref{Eq:dH}) differ for the three surface cells only in the definition of $\Delta E_y|_s$ and $\Delta H_x|_s$. Defining the fields for the surface at the position $(s,l)$ we have for the three cells,

\begin{align*}
	\text{SC/AC:}\quad & \Delta E|_{s,l}^{n+1} = \boxed{E_y|_{k,l}^{n+1}} - \boxed{E_y|_{k-1,l}^{n+1}} \\
	% \text{AC:}\quad & \Delta E|_{s,l}^{n+1} = \boxed{E_y|_{k,l}^{n+1}} - \boxed{E_y|_{k-1,l}^{n+1}} \\
	\text{TAC:}\quad & \Delta E|_{s,l}^{n+1} = \boxed{E_y|_{k,l}^{n+1}} - \boxed{E_y|_{s^-,l}^{n+1}}
\end{align*}

\noindent and

\begin{align*}
	\text{SC:}\quad & \Delta H|_{s,l}^{n+1/2} = \boxed{H_x|_{k+1/2,l}^{n+1/2}} - \boxed{H_x|_{k-1/2,l}^{n+1/2}}\\
	\text{AC:}\quad & \Delta H|_{s,l}^{n+1/2} = \boxed{H_x|_{s^+,l}^{n+1/2}} - \boxed{H_x|_{s^-,l}^{n+1/2}} \\
	\text{TAC:}\quad & \Delta H|_{s,l}^{n+1/2} = \boxed{H_x|_{s^+,l}^{n+1/2}} - \boxed{H_x|_{k-1/2,l}^{n+1/2}},
\end{align*}

\noindent where the difference is obtained by simply using the appropriate nodes on either side of the surface.

\subsection{Surface polarization equations}

The time domain surface polarization equations (\ref{Eq:Pt}) and (\ref{Eq:Mt}) need to be descritized in time. As these equations represent a 2nd order system a trapezoidal formulation was used producing,

\begin{subequations} \label{Eq:Poldt}
\begin{align}
	\left( [\mathbf{C_{p}}] + \frac{\Delta t [\mathbf{G_{p}}]}{2}\right) &\boxed{[\mathbf{P_s}]|_{l}^{n+1}} =  \left( [\mathbf{C_{p}}] - \frac{\Delta t [\mathbf{G_{p}}]}{2}\right) [\mathbf{P_s}]|_{l}^{n}\nonumber\\
	& + \Delta t [\mathbf{F_{p}}] \left( \frac{E_y|_{s,l}^{n+1} + E_y|_{s,l}^n}{2} \right)\\
	\left( [\mathbf{C_{m}}] + \frac{\Delta t [\mathbf{G_{m}}]}{2}\right) &\boxed{[\mathbf{M_s}]|_{l}^{n+1/2}} =  \left( [\mathbf{C_{m}}]- \frac{\Delta t [\mathbf{G_{m}}]}{2} \right) [\mathbf{M_s}]|_{l}^{n-1/2}\nonumber \\
	& + \Delta t [\mathbf{F_{m}}] \left(  \frac{H_x|_{s,l}^{n+1} + H_x|_{s,l}^n}{2} \right).
\end{align}
\end{subequations}

\noindent For the three cells these two equations only differ in the nature of the forcing terms $E_y|_{s,l}$ and $H_x|_{s,l}$ -- the average field at the surface. For the three cells, we have by inspection,

\begin{subequations}
\begin{align}
	\text{SC/AC:}\quad & E_y|_{s,l} = \frac{E_y|_{k-1,l} +  E_y|_{k,l}}{2}\label{Eq:FAC}\\ 
	% \text{AC:}\quad & E_y|_{s,l}^{n+1} =  \frac{E_y|_{k-1,l}^{n} +  E_y|_{k,l}^{n}}{2}\\
	\text{TAC:}\quad & E_y|_{s,l} =  \frac{E_y|_{s^-,l} +  E_y|_{k,l}}{2}
\end{align}
\end{subequations}

\noindent and

\begin{align}
	\text{SC:}\quad & H_x|_{s,l}=  \frac{H_x|_{s^-,l} +  H_x|_{s^+,l}}{2}\notag\\
	\text{AC:}\quad & H_x|_{s,l} =  \frac{H_x|_{k-1/2,l} +  H_x|_{k+1/2,l}}{2}\notag\\
	\text{TAC:}\quad & H_x|_{s,l} =  \frac{H_x|_{k-1/2,l}+  H_x|_{s^+,l}}{2}\label{Eq:ASymForcing}.
\end{align}

\noindent It is important to note that for the two asymmetrical cells the fields applied to the surface $E_y|_{s,l}$ and $H_x|_{s,l}$ are not co-incident in space. For the AC cell the $E_y|_{s,l}$ is applied at the position $k$-1/2 and $H_x|_{s,l}$ at $k$. For the TAC cell $E_y|_{s,l}$ is applied at the position $k$-1/8 and $H_x|_{s,l}$ at $k$-3/8. This mismatch can be expected to produce some error in the simulations. It is largest for the AC cell and a possible modification is to use,

\begin{align}
	\text{AC:}\quad & H_x|_{s,l} =  \frac{H_x|_{k-3/2,l} +  H_x|_{k+1/2,l}}{2}\label{Eq:SymForcing}
\end{align}

\noindent which brings the forcing function into alignment at $k-1/2$. Unless otherwise noted, the AC cell will use the symmetrical forcing formulation. 

\subsection{Special Cell Update Equations}

In addition to equations defining the GSTCs and the polarizations, each cell has a number of nearby nodes which, although placed in the bulk, are dependent on the surface equations and thus must be solved at the same time.

\begin{enumerate}[(a)]

\item For the SC, we have special update equations for the nodes $E_y|_{k-1,l}^{n+1}$ and $E_y|_{k,l}^{n+1}$, given by

\begin{subequations}
		\label{Eq:SCsp}
\begin{align}	
 	&\boxed{E_y|_{k-1,l}^{n+1}}  =  E_y|_{k-1,j}^{n} + \nonumber\\ 
	&\quad \frac{\Delta t}{\epsilon_0}\left(\frac{\boxed{H_x|_{s^-,l}^{n+1/2}} - H_x|_{k-3/2,l}^{n+1/2}}{\Delta} - \frac{\Delta H_z|_{k-1,l}^{n+1/2}}{\Delta}\right) \\
 	&\boxed{E_y|_{k,l}^{n+1}} =  E_y|_{k,j}^{n} + \nonumber\\ 
	&\quad \frac{\Delta t}{\epsilon_0}\left(\frac{H_x|_{k+1/2,l}^{n+1/2}- \boxed{H_x|_{s^+,l}^{n+1/2}}}{\Delta} - \frac{\Delta H_z|_{k,l}^{n+1/2}}{\Delta}\right)
\end{align}
\end{subequations}

\noindent where $\Delta H_z|_{k,l}^{n+1/2} = H_z|_{k,l+1/2}^{n+1/2} - H_z|_{k,l-1/2}^{n+1/2}$.

\item The AC cell only requires one special update for $E_y|_{k-1,l}^{n+1}$, given by

\begin{align}	
	\label{Eq:ACsp}
 	&\boxed{E_y|_{k-1,l}^{n+1}} =  E_y|_{k-1,j}^{n} + \nonumber\\ 
	&\frac{\Delta t}{\epsilon_0}\left(\frac{\boxed{H_x|_{k-1/2,l}^{n+1/2}} - H_x|_{k-3/2,l}^{n+1/2}}{\Delta} - \frac{\Delta H_z|_{k-1,l}^{n+1/2}}{\Delta}\right).
\end{align}

\item The TAC cell has two update equations, one for $E_y|_{k,l}^{n+1}$ and one for $H_x|_{k-1/2,l}^{n+1/2}$, given by

\begin{subequations}
	\label{Eq:TACsp}
\begin{align}	
 	&\boxed{E_y|_{k,l}^{n+1}} =  E_y|_{k,j}^{n} + \nonumber\\ 
	&\quad \frac{\Delta t}{\epsilon_0}\left(\frac{H_x|_{k+1/2,l}^{n+1/2}- \boxed{H_x|_{s^+,l}^{n+1/2}}}{0.75 \Delta} - \frac{\Delta H_z|_{k,l}^{n+1/2}}{\Delta}\right)\\
 	&\boxed{H_x|_{k-1/2,l}^{n+1/2}} =  H_x|_{k-1/2,j}^{n} + \nonumber\\
	&\quad \frac{\Delta t}{\epsilon_0}\left(\frac{E_y|_{s^-,l}^{n+1/2}- \boxed{E_y|_{k-1,l}^{n+1/2}}}{0.75 \Delta}\right).
\end{align}
\end{subequations}

\end{enumerate}
	
\subsection{Implementation}

For each cell configuration, the equations (\ref{Eq:dE}-\ref{Eq:Poldt}) and one of (\ref{Eq:SCsp}), (\ref{Eq:ACsp}) or (\ref{Eq:TACsp}) form a complete set of linear equation defining the surface variables. These equations can be assembled into a compact matrix form, given by

\begin{equation}
[\mathbf{\Gamma_s}] [\mathbf{X_s}] = [\mathbf{F_s}].\label{Eq:FieldMatrix}
\end{equation}

\noindent For the specific case of the SC cell, $[\mathbf{\Gamma_s}]$ and $[\mathbf{F_s}]$ are presented in the appendix as an example. A field evolution is then simply obtained by stepping in time using the following procedure:

\begin{enumerate}[(a)]
	\item Update the bulk H's at $n+1/2$ ($t_{n+1/2}$= $t_n$ + $\Delta t/2$) using previous E's and the boundary conditions.
	\item Update bulk E's at $n+1$ ($t_{n+1}$ = $t_n$ + $\Delta t$) using the H's calculated at $n+1/2$.
	\item Update surface $H_s$'s, and $M_s$ at $n+1/2$ ($t_{n+1/2}$= $t_n$ + $\Delta t/2$) and $E_s$'s and $P_s$s at $n+1$ ($t_{n+1}$ = $t_n$ + $\Delta t$) using $[\mathbf{X_s}] = [\mathbf{\Gamma_s}]^{-1} [\mathbf{F_s}]$.
\end{enumerate}

\section{Validation}

To test the various Yee-cell configurations, the algorithm above was integrated into a standard 1D/2D Yee-cell based simulator, with a metasurface configured as an SC, AC or TAC cell. The simulation setup was configured as a source applied to the left side of the computation region, and perfectly matched layers (PMLs) on the other three sides. A metasurface was placed vertically half way along the $z$-axis, at $z=0$. The intent of this section is to evaluate the robustness and accuracy of the three cell configurations, and do a detailed comparison between them. 

%Initially pulse propagation normally incident to the surface is investigated for transparent and physical surfaces. Then non-normal incident waves (a diverging Gaussian beam) are simulated. The second section looks at the effect of spatial step size on the surface cell accuracy. Finally, the stability of the method with respect to the choice of susceptibilities is presented. 

%\subsection{Pulse Propagation}

\subsection{Transparent Surface}

\begin{figure}[htbp]
\begin{center}
\begin{subfigure}[]{
\psfrag{A}[c][c][0.8]{E-Field, $E_r(t)$}
\psfrag{B}[c][c][0.8]{E-Field, $E_t(t)$}
\psfrag{C}[c][c][0.8]{$z~\mu$m}
\psfrag{D}[l][c][0.6]{Eq.~\eqref{Eq:FT}}
\psfrag{E}[l][c][0.6]{SC}
\psfrag{F}[l][c][0.6]{AC}
\psfrag{G}[l][c][0.6]{TAC}
\includegraphics[width=0.8\columnwidth]{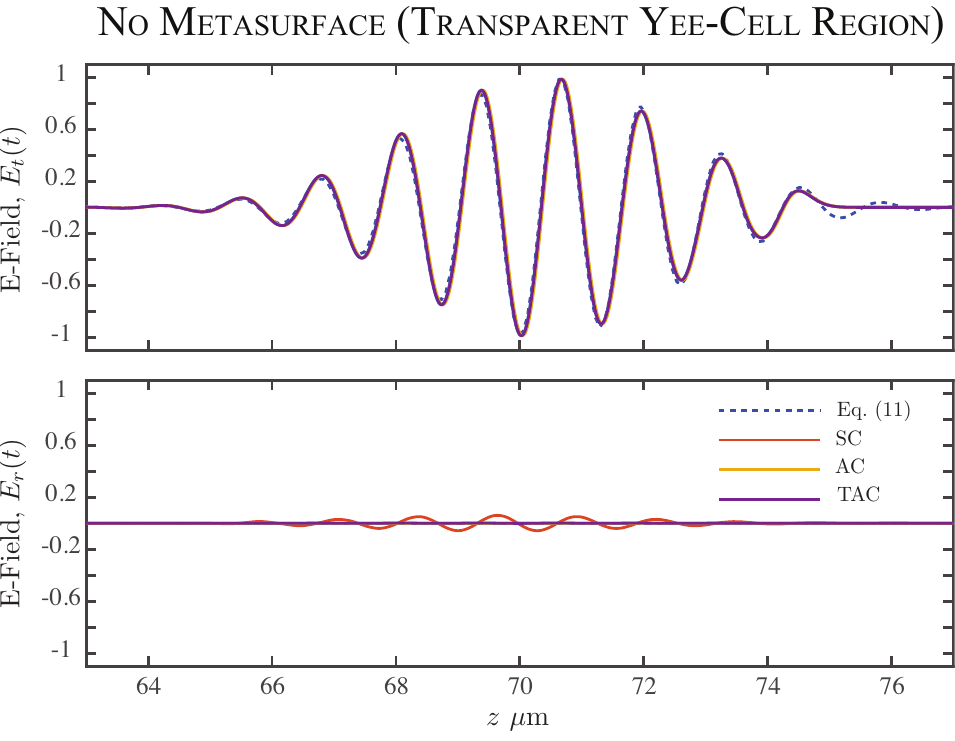}}
\end{subfigure}	
\begin{subfigure}[]{
\psfrag{D}[c][c][0.7]{$z~\mu$m}
\psfrag{B}[c][c][0.7]{$x~\mu$m}
\psfrag{Z}[c][c][0.55]{$\tilde{\chi}_\text{ee} = \tilde{\chi}_\text{mm} = 0$}
\includegraphics[width=\columnwidth]{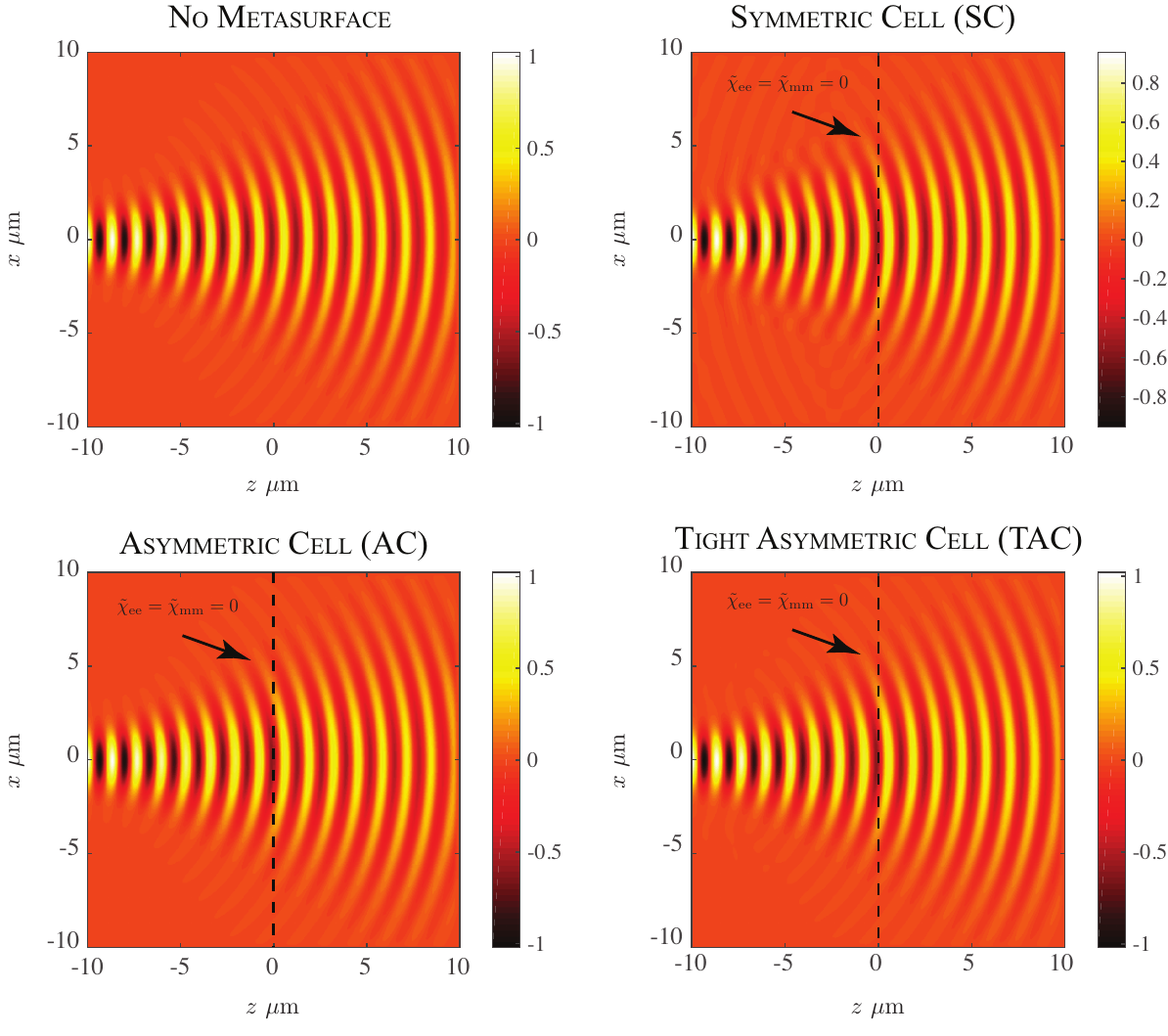}}
\end{subfigure}
\caption{Scattered E-field distribution in the absence of the metasurface (i.e. transparent region), emulated using $\tilde{\chi}_\text{ee} = \tilde{\chi}_\text{mm}=0$, for sanity-check purposes. a) Uniform metasurface excited with a normally incident Gaussian-pulsed plane-wave, compared with analytical Fourier propagation method. b) Steady-state scattered fields for a stepped Gaussian beam. In all cases, fields obtained from all three Yee-cell configurations of Fig.~\ref{Fig:YeeCell} are shown for comparison. The simulations parameters are: Excitation frequency $f_0=230$~THz, Gaussian pulse width $\sigma_t= 10^{-15}$~s, Gaussian beam width $\sigma_x = \lambda_0$,  Yee-cell step size $\Delta = \lambda_0/400$.}\label{Fig:transparent} 
\end{center}
\end{figure}

To provide an initial evaluation of the surface cells, simulations were performed for transparent surfaces (or no metasurface), where $P_s = M_s = 0$. For such a case the GSTCs reduce to 

\begin{equation}
\label{Eq:trans}
	\Delta E_y|_s^{n+1} = 0, \quad \text{and} \quad \Delta H_x|_s^{n+1/2} = 0
\end{equation}

\noindent which gives,

\begin{align}
 \text{SC:}\; &E_y|_{k,l}^{n+1} = E_y|_{k-1,l}^{n+1},~H_x|_{s^+,l}^{n+1/2} = H_x|_{s^-,l}^{n+1/2} \equiv H_x|_{k-1/2,l}^{n+1/2}\notag\\
\text{AC:}\;&E_y|_{k,l}^{n+1} = E_y|_{k-1,l}^{n+1},~H_x|_{k+1/2,l}^{n+1/2} = H_x|_{k-1/2,l}^{n+1/2}\notag\\
\text{TAC:}\; &E_y|_{k,l}^{n+1} = E_y|_{s^-,l}^{n+1},~H_x|_{s^+,l}^{n+1/2} = H_x|_{k-1/2,l}^{n+1/2}.\notag
\end{align}

\noindent As of course, the polarization equations are no longer relevant for a transparent region with no metasurface, which can now be represented by one of the update equations from (\ref{Eq:SCsp}-\ref{Eq:TACsp}) and equation (\ref{Eq:trans}). More specifically, they are given by following for each of the three Yee-cell configurations.

\begin{enumerate}[(a)]

\item For the SC cell, we obtain,

\begin{subequations}
		\label{eq:sctr}
\begin{align}	
 	&E_y|_{k-1,l}^{n+1}  =  E_y|_{k-1,j}^{n} + \\ 
	&\quad \frac{\Delta t}{\epsilon_0}\left(\frac{H_x|_{k-1/2,l}^{n+1/2} - H_x|_{k-3/2,l}^{n+1/2}}{\Delta} - \frac{\Delta H_z|_{k-1,l}^{n+1/2}}{\Delta}\right)\nonumber\\
 	&E_y|_{k,l}^{n+1} =  E_y|_{k,j}^{n} + \\ 
	&\quad \frac{\Delta t}{\epsilon_0}\left(\frac{H_x|_{k+1/2,l}^{n+1/2}- H_x|_{k-1/2,l}^{n+1/2}}{\Delta} - \frac{\Delta H_z|_{k,l}^{n+1/2}}{\Delta}\right)\nonumber
\end{align}
\end{subequations}
\noindent These are identical to bulk equations in absence of a metasurface, however, the relationship $E_y|_{k,l}^{n+1} = E_y|_{k-1,l}^{n+1}$ imposed by the GSTCs (\ref{Eq:trans}) is not correct for this configuration, and we can thus expect some errors to be present.

\item For the AC cell we get,

\begin{align*}	
	\label{Eq:ACsp}
 	&E_y|_{k-1,l}^{n+1} =  E_y|_{k-1,j}^{n} + \nonumber\\ 
	&\frac{\Delta t}{\epsilon_0}\left(\frac{H_x|_{k-1/2,l}^{n+1/2} - H_x|_{k-3/2,l}^{n+1/2}}{\Delta} - \frac{\Delta H_z|_{k-1,l}^{n+1/2}}{\Delta}\right).
\end{align*}

\noindent This is again the bulk update equation. Unlike the SC case, the relationship $E_y|_{k,l}^{n+1} = E_y|_{k-1,l}^{n+1}$ produces the correct update equation this time for the adjacent nodes. The surface is essentially removed from the simulation mesh producing the bulk update equations and we can expect perfect transparency.

\item For the TAC cell, we obtain the following update equations,

\begin{align*}	
 	&E_y|_{k,l}^{n+1} =  E_y|_{k,j}^{n} + \nonumber\\ 
	&\quad \frac{\Delta t}{\epsilon_0}\left(\frac{H_x|_{k+1/2,l}^{n+1/2}- H_x|_{k-1/2,l}^{n+1/2}}{0.75 \Delta} - \frac{\Delta H_z|_{k,l}^{n+1/2}}{\Delta}\right)\\
 	&H_x|_{k-1/2,l}^{n+1/2} =  H_x|_{k-1/2,j}^{n} + \nonumber\\
	&\quad \frac{\Delta t}{\epsilon_0}\left(\frac{E_y|_{k,l}^{n+1/2}- E_y|_{k-1,l}^{n+1/2}}{0.75 \Delta}\right)	.
\end{align*}

\noindent These are identical to the bulk update equations, except for a slight distortion of the cell due to the $0.75$ factor. We can expect near perfect transparency from this configuration.

\end{enumerate}

Figure~\ref{Fig:transparent}(a) shows the time-domain fields in the reflection ($z<0$) and transmission region $(z>0)$, when metasurface is numerically removed by imposing $\tilde{\chi}_\text{ee}(\omega) = \tilde{\chi}_\text{mm}(\omega) = 0$. The figure shows the response of all three Yee-cell configurations (SC, AC and TAC), and compared with analytical Fourier propagation method of Sec.~II-C. The modulation frequency of the input pulse is $230$~THz throughout this paper. The transmitted pulse shows a similar response for all Yee-cell simulations with a slight discrepancy from the Fourier propagation result. The spatial step size was $\Delta = \lambda_0/100$ and was chosen to produce a negligible amount of numerical dispersion. As this dispersion is the same for all cases, this slight pulse distortion can can be attributed to bulk effects, which can be reduced by lowering the spatial step size. The reflection, of course, should ideally be zero, and is found to be negligible for AC and TAC cases. However, the SC exhibits a significant amount of reflection, due to not producing the bulk update equations (see (\ref{eq:sctr})). 

Next, a stepped continuous-wave (CW) Gaussian beam with a width of $\lambda_0$ was launched from the left side, with strongly divergent wavefronts. Fig. \ref{Fig:transparent}(b) shows the computed steady-state E-field distribution obtained for each of the three Yee-cell configurations, and compared with analytical Fourier propagation method. As can be seen, no significant reflection is present except in the case of a SC configuration. We can conclude from these observations, that for transparent or nearly transparent surfaces (i.e, $\chi_\text{ee, mm} \approx 0$), the SC is inappropriate due existence of spurious numerical reflections. 

\begin{figure}[htbp]
\begin{center}
\begin{subfigure}[]{
\psfrag{A}[c][c][0.8]{E-field, $E_r$}
\psfrag{B}[c][c][0.8]{E-Field, $E_t$}
\psfrag{C}[c][c][0.8]{$z~\mu$m}
\psfrag{D}[c][c][0.8]{$\boxed{\Delta = \lambda_0/25}$}
\psfrag{H}[c][c][0.8]{$\boxed{\Delta = \lambda_0/100}$}
\psfrag{F}[c][c][0.8]{$\boxed{\Delta = \lambda_0/400}$}
\psfrag{d}[l][c][0.6]{Eq.~\eqref{Eq:FT}}
\psfrag{e}[l][c][0.6]{SC}
\psfrag{f}[l][c][0.6]{AC}
\psfrag{g}[l][c][0.6]{TAC}
\psfrag{E}[l][c][0.8]{$\tilde{\chi}_\text{ee}(\omega) = \tilde{\chi}_\text{mm}(\omega)$}
\includegraphics[width=0.75\columnwidth]{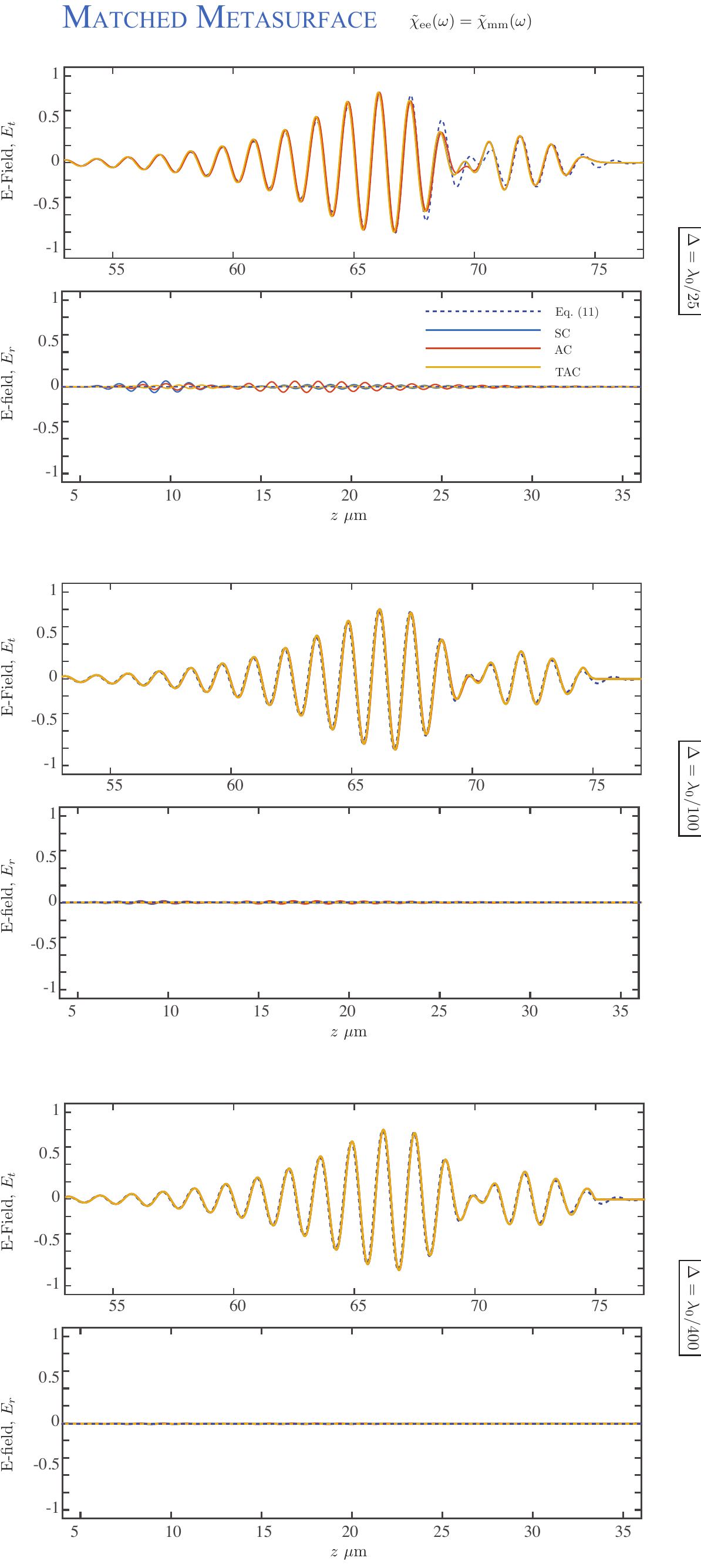}}
\end{subfigure}	
\begin{subfigure}[]{
\psfrag{A}[c][c][0.8]{E-field, $E_r$}
\psfrag{B}[c][c][0.8]{E-field, $E_t$}
\psfrag{C}[c][c][0.8]{$z~\mu$m}
\psfrag{F}[l][c][0.6]{Eq.~\eqref{Eq:FT}}
\psfrag{D}[c][c][0.8]{$\boxed{\Delta = \lambda_0/400}$}
\psfrag{E}[l][c][0.8]{$\tilde{\chi}_\text{ee}(\omega) \ne \tilde{\chi}_\text{mm}(\omega)$}
\includegraphics[width=0.75\linewidth]{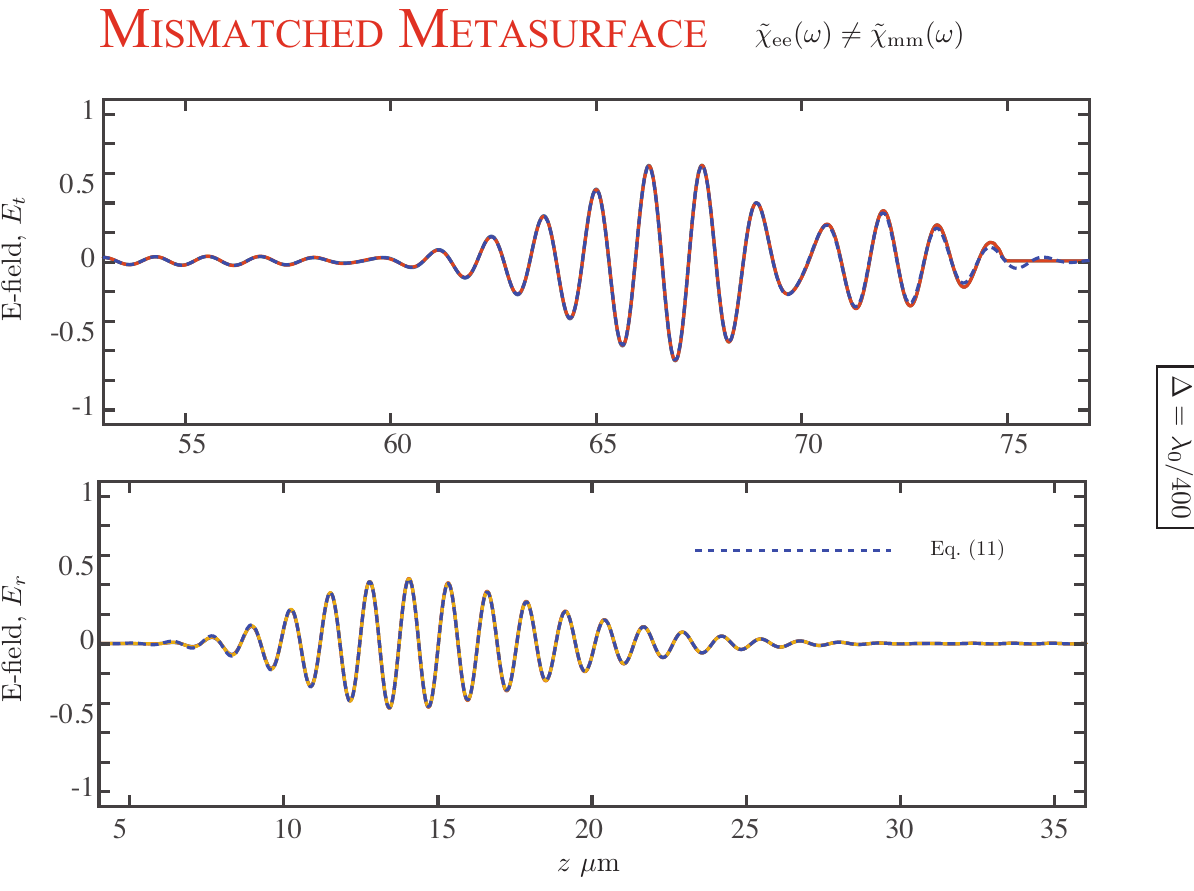}}
\end{subfigure}	
\caption{Transmission and reflection response of a uniform metasurface excited with a normally incident Gaussian-pulsed plane-wave, corresponding to a) a matched metasurface [$\tilde{\chi}_\text{ee}(\omega) = \tilde{\chi}_\text{mm}(\omega)$] with gradually increasing spatial step-size $\Delta$, and b) mis-matched metasurface [$\tilde{\chi}_\text{ee}(\omega) \ne \tilde{\chi}_\text{mm}(\omega)$]. The Gaussian pulse parameters are the same as in Fig.~\ref{Fig:transparent}. The metasurface susceptibilities are defined using a single Lorentzian, for simplicity, with the following parameters: $\omega_{ep} = 3.01\times 10^{11}$~rad/s, $\omega_{e0}=2\pi(230~\text{THz})$ and $\gamma_e = 7.54\times10^{12}$. For the mismatched metasurfaces $\omega_{e0} - \omega_{m0} =2\pi(15~\text{GHz})$. The input pulse is Gaussian with full-width-half-maximum (FWHM) of $10^{-15}$~s with a modulation frequency of $230$~THz. }\label{Fig:PulsedResults} 
\end{center}
\end{figure}

\begin{figure}[htbp]
\begin{center}
\psfrag{A}[c][c][0.8]{E-field, $E_r$}
\psfrag{B}[c][c][0.8]{E-field, $E_t$}
\psfrag{C}[c][c][0.8]{$z~\mu$m}
\psfrag{D}[l][c][0.6]{Eq.~\eqref{Eq:FT}}
\psfrag{E}[l][c][0.6]{Symmetrical Forcing Function, \eqref{Eq:SymForcing}}
\psfrag{F}[l][c][0.6]{Asymmetrical Forcing Function, \eqref{Eq:ASymForcing}}
\includegraphics[width=0.8\linewidth]{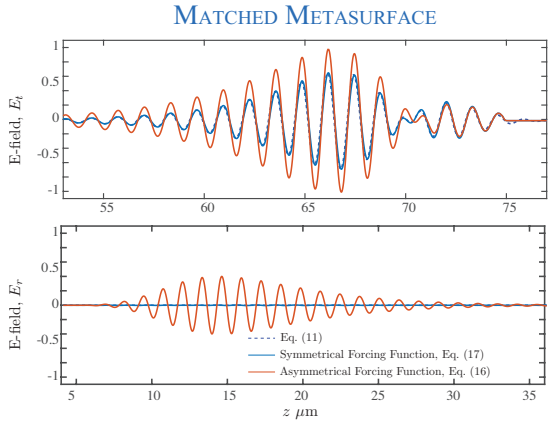}
\caption{Comparison of the transmitted and reflected fields from a uniform metasurface of Fig.~\ref{Fig:PulsedResults}, computed using the asymmetric cell (AC), using a symmetric and asymmetric forcing function in the surface polarization.}\label{Fig:ForcingFunctionAC} 
\end{center}
\end{figure}

\subsection{Lorentzian Metasurface}

Let us now introduce a metasurface inside the computational region, at $z=0$. Fig.~\ref{Fig:PulsedResults}(a) shows the transmitted and reflected pulses, corresponding to a normally incident Gaussian pulsed plane-wave on a matched surface ($\chi_\text{ee} = \chi_\text{mm}$) for three different step sizes ($\Delta = \lambda_0/25, \lambda_0/100$ and $\lambda_0/400$) and a symmetric forcing function in the surface susceptibilities. As can be seen all three surfaces produce similar results. The pulse is strongly dispersed by the presence of a metasurface with Lorentzian susceptibilities. Furthermore, the metasurface response matches very well with the FT propagation result albeit with some numerical dispersion determined by the spatial step size. A matched metasurface should produce no reflections, and the FT result does show this. For the larger step sizes all of the surfaces produce non-negligible reflection with the AC cell, arguably producing the most and the TAC the least. For all the Yee-cell configurations, the field reflection can be reduced to a negligible value by decreasing the spatial step size to $\Delta = \lambda_0/400$. Figure~\ref{Fig:PulsedResults}(b) further shows the transmitted and reflected fields for an unmatched metasurface ($\chi_\text{ee} \ne \chi_\text{mm}$) with a spatial step size of $\Delta = \lambda_0/400$, where a significant amount of reflection is expected. All of the Yee-cell configurations produce an excellent match to the FT result, with again the TAC cell producing the least amount of error. Finally, to further evaluate the impact of the forcing function in the Lorentzian susceptibilities, a simulation using the AC cell with asymmetrical forcing (\ref{Eq:ASymForcing}) is also shown in Fig. \ref{Fig:ForcingFunctionAC}. It can be clearly seen that even for a very small step size ($\Delta=\lambda_0/400$), significant reflections are produced. This concludes that the Lorentzian polarizabilities must be excited with symmetrical forcing functions for all three Yee-cell configurations.

\begin{figure}[htbp]
\begin{center}
\begin{subfigure}[]{
\psfrag{A}[c][c][0.8]{Transmitted Energy, $|E_t/E_0|^2$}
\psfrag{B}[c][c][0.8]{Steps per wavelength $\lambda_0/\Delta$}
\psfrag{C}[c][c][0.8]{Reflected Energy, $|E_r/E_0|^2$}
\psfrag{G}[l][c][0.6]{Eq.~\eqref{Eq:FT}}
\psfrag{E}[l][c][0.6]{SC}
\psfrag{D}[l][c][0.6]{AC}
\psfrag{F}[l][c][0.6]{TAC}
\includegraphics[width=0.75\linewidth]{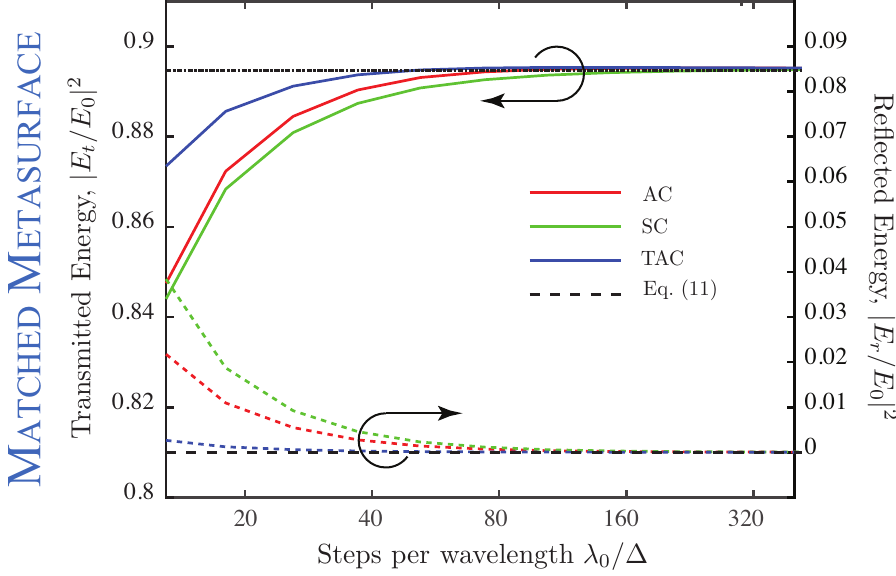}}
\end{subfigure}	
\begin{subfigure}[]{
\psfrag{A}[c][c][0.8]{Transmitted Energy, $|E_t/E_r|^2$}
\psfrag{B}[c][c][0.8]{Steps per wavelength $\lambda_0/\Delta$}
\psfrag{C}[c][c][0.8]{Reflected Energy, $|E_r/E_r|^2$}
\psfrag{G}[l][c][0.6]{Eq.~\eqref{Eq:FT}}
\psfrag{E}[l][c][0.6]{SC}
\psfrag{D}[l][c][0.6]{AC}
\psfrag{F}[l][c][0.6]{TAC}
\includegraphics[width=0.75\linewidth]{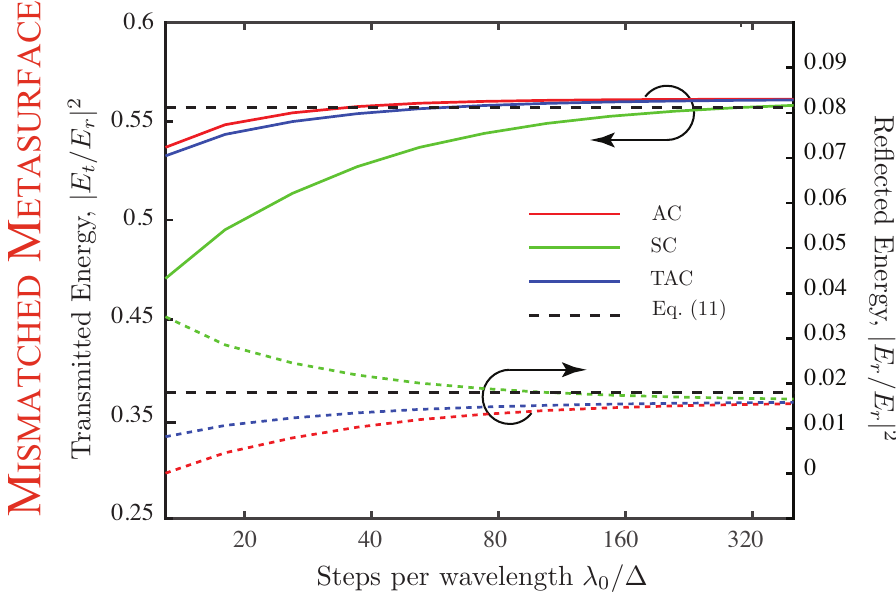}}
\end{subfigure}	
\caption{The convergence plots for the transmitted and reflected fields of Fig.~\ref{Fig:PulsedResults}, as a function of Yee-cell grid spatial step size, for the cases of a) matched metasurface and b) mismatched metasurface, corresponding to all three Yee-cell configurations of Fig.~\ref{Fig:YeeCell}, and compared to analytical Fourier propagation results.}\label{Fig:Convergence}
\end{center}
\end{figure}

Figure~\ref{Fig:Convergence} further shows the impact of the spatial step size on the scattered fields for both cases of matched and mismatched metasurface. The total normalized EM energy is computed in both transmission and reflections regions as a function of step size $\Delta$, and compared with the ideal results obtained using the Fourier transform propagation method. A monotonic convergence is observed for all the three Yee-cell configurations, and in particular, AC and TAC is seen to be converged faster than the SC, in all cases. It should be noted that the total normalized energy in the computation region is less than 1, due to non-zero $\alpha$'s accounting for dissipation losses in the metasurface.

\begin{figure}[tbp]
\begin{center}
\psfrag{D}[c][c][0.7]{$z~\mu$m}
\psfrag{B}[c][c][0.7]{$x~\mu$m}
\includegraphics[width=\columnwidth]{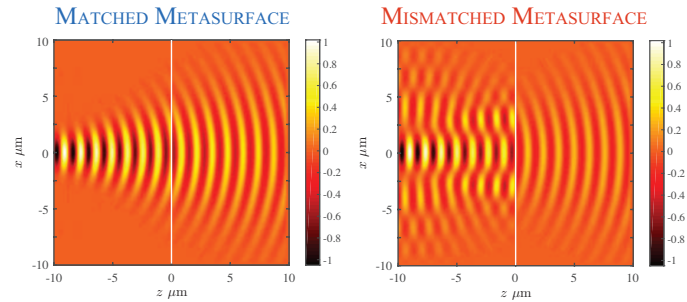}
\caption{Steady-state scattered E-fields of a stepped Gaussian beam input excitation, shown for a matched and mismatched metasurface. All the simulation parameters are the same as in Fig.~\ref{Fig:PulsedResults}. The signal step is modelled using a slowly rising Gaussian edge of FWHM $10^{-5}$~s.}\label{Fig:GaussianMetasurface}
\end{center}
\end{figure}

Finally, Fig.~\ref{Fig:GaussianMetasurface} shows the scattered fields from a matched and mismatched metasurface, excited with a stepped CW Gaussian beam, computed using the TAC configuration. As expected, the matched surface produces a phase discontinuity at the surface but with no disruption of the beam propagation. On the other hand, the mismatched metasurface creates a more complex response as the reflected waves from the surface cause a standing wave to form  in the reflection region, as compared to purely forward propagating waves in the transmission region. 

\section{Causality Considerations for Constant Surface Susceptibilities}

The prime motivation for modelling metasurface susceptibility densities using Lorentzian profiles inside of a Yee-cell region, is to incorporate a physically motivated causal response from the metasurface sub-wavelength unit cells. While an assumption of a constant susceptibility will greatly simplify the proposed Yee-cell algorithm, it fails to capture two important physical effects (under some conditions): a) Causality, and b) Dispersion. While it is clear that constant surface susceptibilities are inherently non-dispersive, the causality aspect is not straightforward. This aspect is further clarified in this section, to emphasize the importance of using Lorentzian susceptibilities in the proposed FDTD method.

Let us consider a lossless uniform matched metasurface [$\tilde{\chi}_\text{ee}(\omega) = \tilde{\chi}_\text{mm}(\omega) = \chi_0$, where $\chi_0\in\mathcal{R}$] excited with a pulsed Gaussian plane-wave, whose envelope is given by $E_0(t, z=0_-) = \exp[-(t/T_0)^2]$. The transmission function of a matched metasurface is given by \eqref{Eq:TDGSTCF}, as

\begin{equation}
T(\omega) = \left(\frac{2c - j\omega \chi_0}{2c + j\omega \chi_0}\right) = -\exp\left\{2\tan^{-1}\left(\frac{\omega\chi_0}{2c} \right\}\right),
\end{equation}

\noindent where the last equality is based on the fact that $|T(\omega)|=1~\forall~\omega$, i.e. an all-pass transfer function. For small arguments of the inverse tangent function, $T(\omega) \approx \exp\{-j \omega \chi_0/c\}$, so that the output of the metasurface is given by

\begin{equation}
E_t(t, 0_+) = \mathcal{F}^{-1}\{\mathcal{F}[E_0(t, 0_-)] T(\omega)\}= \exp\left[-\left(\frac{t - k_0\chi_0}{T_0}\right)^2\right]\notag,
\end{equation}

\noindent where $k_0=\omega/c$. Therefore, the metasurface output is also a Gaussian pulse, as expected, however its peak is now located at a time instant $t_\text{peak} = k_0\chi_0$. There are now two following possibilities:

\begin{enumerate}
\item $\chi_0>0$: The output pulse is located at $t_\text{peak} \ge 0$, i.e. a positive time-delay.
\item $\chi_0<0$: The output pulse is now located at $t_\text{peak} < 0$, i.e. a negative time-delay or a time advance.
\end{enumerate}

\noindent While Case 1 is naturally causal, Case 2 represents a non-causal response, as the output pulse appears before the input. Therefore, for this simple case of a matched metasurface, it can be concluded that \emph{negative and constant surface susceptibilities}, represent a non-physical system, and thus is not allowed. This is consistent with the fact that causal EM metamaterials with negative constitutive parameters $\epsilon<0$ and $\mu<0$, must be \emph{dispersive} to allow positive time-average stored electric and magnetic energies \cite{Rothwell}\cite{Caloz_Wiley_2006}.  

\begin{figure}[tbp]
\centering
\psfrag{A}[c][c][0.75]{E-field, $E_t$}
\psfrag{B}[c][c][0.75]{E-field, $E_t$}
\psfrag{c}[c][c][0.75]{$z~\mu$m}
\psfrag{D}[l][c][0.6]{input $E_0$}
\psfrag{E}[l][c][0.6]{Eq.~\eqref{Eq:FT}}
\psfrag{F}[l][c][0.6]{TAC simulation}
\psfrag{G}[c][c][0.6]{$+$tive delay}
\psfrag{H}[c][c][0.6]{$-$tive delay}
\psfrag{K}[c][c][0.6]{$\text{Re}\{\tilde{\chi}_\text{ee}(\omega)\} > 0$}
\psfrag{M}[c][c][0.6]{$\text{Re}\{\tilde{\chi}_\text{ee}(\omega)\} < 0$}
\psfrag{J}[c][c][0.6]{\shortstack{numerical \\instability}}
\includegraphics[width=\columnwidth]{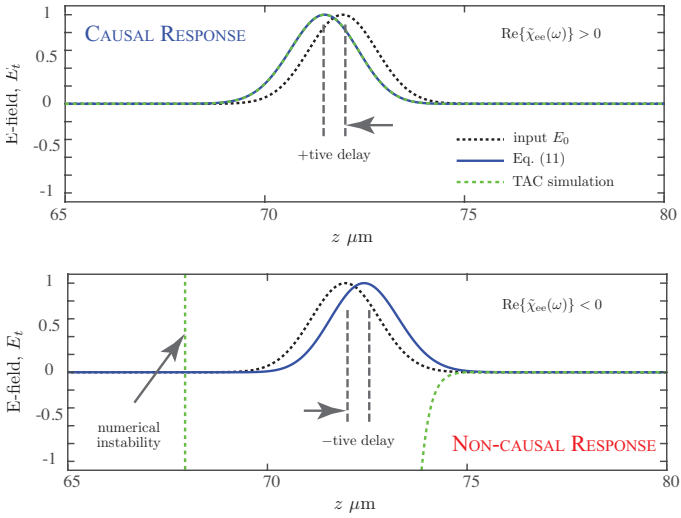}\\
\caption{Demonstration of a non-causal response from the metasurface (assumed to be matched, for simplicity), whose surface susceptibilities are defined as frequency independent. a) $\text{Re}\{\tilde{\chi}_\text{ee}(\omega)\} \ge 0~\forall~\omega$, and b) $\text{Re}\{\tilde{\chi}_\text{ee}(\omega)\} < 0~\forall~\omega$. Simulation parameters are $\tilde{\chi} = 4.8731\times 10^{-7} - j2.8616\times 10^{-8}. $b) $\tilde{\chi} = -5.0881\times 10^{-7} -j 3.1207\times 10^{-8}$. }\label{Fig:ConstantChi}
\end{figure}

To numerically demonstrate this, consider a broadband pulsed excitation of a uniform metasurface with a very short Gaussian pulse. Fig.~\ref{Fig:ConstantChi} shows the computed response corresponding to both analytical Fourier transform propagation approach and the proposed Yee-cell using the TAC configuration, for the two cases when $\text{Re}\{\chi_0\}>0$ and $\text{Re}\{\chi_0\}<0$. As expected, the TAC simulation and the FT simulation match perfectly for the first case providing a positive time delay. The second case with negative real part of $\chi_0$ produces a phase advance for the transmitted field, thereby creating a non-causal response in the FT simulation with the transmitted pulse running ahead of the excitation envelope. As we would expect, for the Yee-cell simulation, this causes a numerical instability which is indicated by the amplification and gross distortion of the transmitted pulse \cite{NakhlaCS}. This further demonstrates the need for the use of a physical causal surface representation, such as a Lorentzian response in a numerical Yee-cell model.

\section{Conclusions}

An FDTD simulation of broadband electromagnetic metasurfaces has been proposed based on direct incorporation of GSTCs inside of Yee-cell region, for arbitrary wave excitations. This has been achieved by inserting a zero thickness metasurface inside bulk nodes of the Yee-cell region, giving rise to three distinct cell configurations - SC, AC and TAC. In addition, the metasurface has been modelled using electric and magnetic surface susceptibilities exhibiting a broadband Lorentzian response. As a result, the proposed model guarantees a physical and causal response from the metasurface. Several full-wave results have been shown, and compared with analytical Fourier propagation methods showing excellent results, for both 1D and 2D fields simulations. It has been further found that the TAC provides the fastest convergence among the three methods with minimum error. While only scalar surface susceptibilities exhibiting a single Lorentzian resonant contribution is described here for simplicity, the proposed method can be easily extended to full tensorial susceptibilities with arbitrary number of Lorentzians, in a fully 3D simulation environment.

\section*{Appendix}

\subsection{Symmetrical Surface Cell Matrix Formulation}

To illustrate the formation of field matrix equation \eqref{Eq:FieldMatrix}, let us take an example of a symmetric cell among the possible three configurations of Fig.~\ref{Fig:YeeCell}. The equations describing the symmetric cell are given by (\ref{Eq:dE}-\ref{Eq:Poldt}) and (\ref{Eq:SCsp}). Using them, we obtain the following set of field equations:

\begin{figure*}[htbp]
\begin{align}
	[\mathbf{X_{SC}} & = \left[E_y|_{k-1,l}^{n+1} \; E_y|_{k,l}^{n+1}  \;  H_x|_{s^-,l}^{n+1/2} \; H_x|_{s^+,l}^{n+1/2} \; [\mathbf{P_s}]|_{l}^{n+1/2} \; [\mathbf{M_s}]|_{l}^{n+1/2} \right]^T \label{Eq:Mat1}\\
 	[\mathbf{\Gamma_{SC}}]  &=
 	\left[
 		\begin{array}{cccccc}
 			{-\frac{\Delta t}{\mu_0}} & {\frac{\Delta t}{\mu_0}} & 0 & 0   & [ 0 \quad 1 ] & [ 0 \quad 0 ] \\
 			0 & 0 & -\frac{\Delta t}{\epsilon_0} & \frac{\Delta t}{\epsilon_0} & [ 0 \quad 0 ] & [ 0 \quad 0 ] ]\\
 			1 & 0 & -\frac{\Delta t}{\epsilon_0 \Delta} & 0 & [ 0 \quad 0 ] & [ 0 \quad 1 ]\\
 			0 & 1 & 0 & \frac{\Delta t}{\epsilon_0 \Delta}  & [ 0 \quad 0 ] & [ 0 \quad 0 ]\\
  			\frac{-\Delta t [\mathbf{F_p}]}{4}& \frac{-\Delta t [\mathbf{F_p}]}{4}  & 0 & 0 & \left( [\mathbf{C_p}] + \frac{\Delta t [\mathbf{G_p}]}{2}\right) & \varnothing  \\
			0 & 0 & \frac{-\Delta t [\mathbf{F_m}]}{4} & \frac{-\Delta t [\mathbf{F_m}]]}{4}  & \varnothing & \left( [\mathbf{C_m}] + \frac{\Delta t [\mathbf{G_m}]}{2}\right) \\
 	  	\end{array}
 	\right]\label{Eq:Mat2} \\
 	[\mathbf{ F_{SC}} & =
 	\left[
 		\begin{array}{c}
 			(  - \mu_0 \frac{M_s|_l^{n-1/2}}{\Delta t})\frac{\Delta t}{\mu_0} \\
 			(  - \epsilon_0\frac{Q_s|_l^{n-1}}{\Delta t})\frac{\Delta t}{\epsilon_0}\\
 			E_y|_{k-1,l}^{n} +   \frac{\Delta t}{\epsilon_0 \Delta}(-H_x|_{k-3/2,l}^{n+1/2} - H_z|_{k-1,l+1/2}^{n+1/2} + H_z|_{k-1,l-1/2}^{n+1/2})\\
 			E_y|_{k,l}^{n} +   \frac{\Delta t}{\epsilon_0 \Delta}(H_x|_{k+1/2,l}^{n+1/2} - H_z|_{k,l+1/2}^{n+1/2} + H_z|_{k,l-1/2}^{n+1/2})\\
 			([\mathbf{C_p}] - \frac{\Delta t [\mathbf{G_p}]}{2}) [\mathbf{Q_s}]|_{l}^{n} +  \frac{\Delta t [\mathbf{F_p}]}{4}  (E_y|_{k,l}^{n} + E_y|_{k-1,l}^{n}) \\
 			([\mathbf{C_m} - \frac{\Delta t [\mathbf{G_m}]}{2}) [\mathbf{M_s}]|_{l}^{n-1/2} + \frac{\Delta t [\mathbf{F_m}]}{4}  (H_x|_{s^-,l}^{n-1/2} + H_x|_{s^+,l}^{n-1/2}) \\ 
 	  	\end{array}
	\right]  \label{Eq:Mat3}\\ \notag \\ \hline \notag
\end{align}
\end{figure*}

\noindent 1) The pair of GSTCs for both $E$- and $H$-fields.

\begin{subequations}
\begin{equation}
\boxed{E_y|_{k,l}^{n+1}} - \boxed{E_y|_{k-1,l}^{n+1}} =  \mu_0 \frac{\boxed{M_s|_l^{n+1/2}} - M_s|_l^{n-1/2}}{\Delta t}\notag
\end{equation}
\begin{equation}
\boxed{H_x|_{k+1/2,l}^{n+1/2}} - \boxed{H_x|_{k-1/2,l}^{n+1/2}} = \frac{\boxed{P_s|_{l}^{n+1}} - P_s|_{l}^{n}}{\Delta t}\notag
\end{equation}
\end{subequations}

\noindent 2) The pair of special update equations, for the nodes $E_y|_{k-1,l}^{n+1}$ and $E_y|_{k,l}^{n+1}$.

\begin{align}
\boxed{E_y|_{k-1,l}^{n+1}}  &= E_y|_{k-1,j}^{n} \notag\\
&   + \quad \frac{\Delta t}{\epsilon_0}\left(\frac{\boxed{H_x|_{s^-,l}^{n+1/2}} - H_x|_{k-3/2,l}^{n+1/2}}{\Delta} - \frac{\Delta H_z|_{k-1,l}^{n+1/2}}{\Delta}\right)\notag\\
\boxed{E_y|_{k,l}^{n+1}} &=  E_y|_{k,j}^{n} \notag\\
&+ \quad \frac{\Delta t}{\epsilon_0}\left(\frac{H_x|_{k+1/2,l}^{n+1/2}- \boxed{H_x|_{s^+,l}^{n+1/2}}}{\Delta} - \frac{\Delta H_z|_{k,l}^{n+1/2}}{\Delta}\right)\notag
\end{align}

\noindent 3) The pair of Lorentzian surface polarization densities for both $E-$ and $H$-fields.

\begin{align}
&\left( [\mathbf{C_{p}}] + \frac{\Delta t [\mathbf{G_{p}}]}{2}\right) \boxed{[\mathbf{P_s}]|_{l}^{n+1}} =  \left( [\mathbf{C_{p}}] - \frac{\Delta t [\mathbf{G_{p}}]}{2}\right) [\mathbf{P_s]}|_{l}^{n} \notag\\
&+ \Delta t [\mathbf{F_{p}}] \left( \frac{E_y|_{k-1,l}^{n+1} + E_y|_{k,l}^{n+1} + E_y|_{k-1,l}^n + E_y|_{k,l}^n}{4} \right)\notag\\
&\left( [\mathbf{C_{m}}] + \frac{\Delta t [\mathbf{G_{m}}]}{2}\right) \boxed{[\mathbf{M_s}]|_{l}^{n+1/2}} =  \left( [\mathbf{C_{m}}]- \frac{\Delta t [\mathbf{G_{m}}]}{2} \right) [\mathbf{M_s}]|_{l}^{n-1/2}\nonumber \notag \\
&+ \Delta t [\mathbf{F_{m}}] \left(  \frac{H_x|_{s^-,l}^{n+1} + H_x|_{s^+,l}^{n+1} + H_x|_{s^-,l}^{n} + H_x|_{s^+,l}^n}{4} \right).\notag
\end{align}

\noindent These equations can be placed in an appropriate matrix forms as shown in \eqref{Eq:Mat1}-\eqref{Eq:Mat3}, which can be now used to update the subsequent fields on the Yee-cell nodes.

\bibliographystyle{IEEETran}
\bibliography{Smy_Implicit_FDTD_Metasurface_TAP_2017}

\end{document}